\documentclass[namedreferences,hyperref,optionalrh,solaromanenum]{spr-sola}

\usepackage{graphicx}                    
\usepackage{amssymb}                    
\usepackage{amsmath}
\usepackage{color}                       
\usepackage{breakurl}                         

\newcommand{\V}{\mathcal{V}}
\newcommand{\I}{\mathcal{I}}

\newcommand{\B}{\mathcal{B}}
\newcommand{\rms}{{\sl rms}\rm}

\chardef\us=`\_

\begin{document}
\begin{article}
\begin{opening}

\title{Noise in Maps of the Sun at Radio Wavelengths\\I: Theoretical Considerations}

%
\author[addressref={aff1},corref,email={tbastian@nrao.edu}]{\inits{T. S. }\fnm{Timothy }\snm{Bastian}\orcid{0000-0002-0713-0604}}
\author[addressref={aff2},corref,email={bin.chen@njit.edu}]{\inits{B. }\fnm{Bin }\snm{Chen}\orcid{}}
\author[addressref={aff2},corref,email={surajit.mondal@njit.edu}]{\inits{S. }\fnm{Surajit }\snm{Mondal}\orcid{0000-0002-2325-5298}}
\author[addressref={aff3},corref,email={shilaire@berkeley.edu}]{\inits{P. }\fnm{Pascal }\snm{Saint-Hilaire}\orcid{}}

%
\runningauthor{Bastian et al.}
\runningtitle{Noise in Radio Maps of the Sun}

\address[id={aff1}]{National Radio Astronomy Observatory, 520 Edgemont Rd, Charlottesville, VA 22903, USA}
\address[id={aff2}]{Center for Solar-Terrestrial Research, New Jersey Institute of Technology, 323 M L King Jr Boulevard, Newark, NJ 07102, USA}
\address[id={aff3}]{Space Sciences Laboratory, University of California, 7 Gauss Way, Berkeley, CA, 94720 USA}

\begin{abstract}
The Sun is a powerful source of radio emissions, so much so that, unlike most celestial sources, this emission can dominate the system noise of radio telescopes. We outline the theory of noise in maps formed by Fourier synthesis techniques at radio wavelengths, with a focus on self-noise: that is, noise due to the source itself. As a means of developing intuition we consider noise for the case of a single dish, a two-element interferometer, and an $n$-element array for simple limiting cases. We then turn to the question of the distribution of noise on a map of an arbitrary source observed at radio wavelengths by an $n$-element interferometric array. We consider the implications of self-noise for observations of the Sun in a companion paper. 
\end{abstract}


\end{opening}


\pagebreak

 \section{Introduction}\label{Intro} 

Fourier synthesis mapping at radio wavelengths using an interferometric array of antennas is a mature technique for imaging celestial sources with a high degree of angular resolution over a large field of view. The technique has been exploited by ground based telescopes spanning more than four orders of magnitude in wavelength, from decameter/meter wavelengths (e.g., the LOw Frequency ARray - LOFAR: \citealt{Haarlem2013}) to millimeter/submillimeter wavelengths (e.g., the Atacama Large Millimeter/submillimeter Array - ALMA: \citealt{Wootten2009}). Most Fourier synthesis arrays are optimized for detecting and imaging faint celestial sources in their continuum or spectral line emissions. For the vast majority of sources observed by such arrays, the emission from the source itself makes no significant contribution to the system noise. However, sources for which the source itself contributes significant noise do exist — examples include pulsars \citep{Oslowski2011}, radio galaxies such as Virgo A \citep{Morgan2021}, supernova remnants \citep{McCullough1993}, and the Sun \citep{Bastian1989}. The role of noise contributed by the source cannot necessarily be ignored for these sources. Depending on the nature of the source and the array used to observe it, ``self-noise'' can manifest itself in complex ways across a map. 

In this paper we consider the properties of noise in Fourier synthesis images of strong sources from a theoretical perspective. In a companion paper (Paper II), we consider these ideas for the specific problem of imaging solar phenomena at radio wavelengths. In the next section we begin with a brief introduction to Fourier synthesis mapping. In Section~3 we introduce some terminology and provide an overview of noise properties in radio observations. To develop intuition, we consider simple sources using single-dish observations, a correlating two-element interferometer, and a correlating $n$-element array. In Section~4 we consider the general case of noise properties of synthesis images both on and off radio sources using a correlating array with and without total power measurements, and discuss prospects for mitigating the self-noise contribution. We refer to the former case as a ``correlation array" and the latter as a ``total power" array. We conclude in Section~5. 

\section{Fourier Synthesis Imaging}\label{Fourier}

It is assumed that reader is broadly familiar with interferometry and Fourier synthesis imaging at radio wavelengths. Nevertheless, we begin by introducing some key concepts and providing a brief overview of synthesis imaging at radio wavelengths before discussing noise in radio interferometric measurements and the resulting maps. 

The angular resolution of a single dish at radio wavelengths is $\theta_{\rm SD}\sim \lambda/D$, where $D$ is the diameter of the (circular) antenna, $\lambda$ is the wavelength of the radiation in question. The only way the angular resolution can be increased at a given wavelength using a single dish is to increase $D$, an option that quickly becomes too costly to be a realistic in most cases. In addition, a single dish is limited to observing a single pixel on the sky unless mapping techniques are used: e.g., pointing to a discrete grid on the sky, driving the telescope continuously across the sky in a defined sampling pattern, or the use of a focal plane array. 

The utility of interferometric arrays of antennas operating at radio wavelengths is their ability to image celestial sources with a high degree of angular resolution. The limitations to angular resolution imposed by a single dish are overcome by, in effect, dividing the large single aperture into a number of small apertures and distributing them in some optimum configuration over an effective aperture. The angular resolution of the interferometric array $\theta_{\rm INT}$ is then determined by the maximum separation between antennas $d_{max}$:  $\theta_{\rm INT}\sim \lambda/d_{\rm max}$. The field of view (FOV) of the array is determined by the angular response of the constituent antennas (each presumed here to be identical circular apertures of diameter $D$), referred to as the {\sl primary beam}. The primary beam is often well-approximated by a Gaussian with a full width at half maximum $\theta_{\rm FOV}=\theta_{\rm SD}\sim \lambda/D$ and we have $\theta_{\rm INT}\ll\theta_{\rm FOV}$. The disadvantage to using this technique is that the effective aperture is usually incompletely sampled. 

Consider a source with a radio brightness distribution on the sky at a cyclic frequency $\nu=c/\lambda$ given by $\I(l,m)$, where $l$ and $m$ are direction cosines relative to the phase tracking center of the array, usually the center of the field of view. The units of $\I(l,m)$ are those of specific intensity: W~m$^{-2}$~Hz$^{-1}$~sr$^{-1}$. We will also refer to the {\sl spectral flux density}, or simply ``flux density", the specific intensity integrated over solid angle. In radio astronomy, the unit for spectral flux density is the Jansky, where 1~Jy$=10^{-26}$ W m$^{-2}$ Hz$^{-1}$. Since solar signals have much larger flux densities than other celestial sources, {\sl solar flux units} (SFU) are often used for solar observations, where 1~SFU $\equiv 10^4$~Jy. 

The fundamental relationship between $\I(l,m)$ and measurements made by an interferometric array is based on the van~Cittert-Zernike theorem (e.g., \citealt{Born1980}):

\begin{equation}
\V(u,v)=\int\int A_\circ(l,m) \I(l,m) e^{-2\pi i(ul+vm)} dl dm,
\end{equation}

\noindent where $\V(u,v)$ is referred to as the {\sl visibility} function, a complex quantity. $A_\circ(l,m)$ is the normalized primary beam, and the coordinates $u$ and $v$ represent differences in position of any two antennas $j$ and $k$ expressed in wavelength units: $u=\nu(x_j-x_k)/c$ and $v=\nu(y_j-y_k)/c$. Note that  $l=\sin\theta_x\approx \theta_x$ and $m=\sin\theta_y\approx \theta_y$ for small angular offsets from the phase tracking center which, for our purposes, we take to be ($\theta_x,\theta_y)=(0,0)$. In this case, Equation~1 can be inverted via inverse Fourier transform to yield

\begin{equation}
A_\circ(l,m) \I(l,m)=\int\int \V(u,v) e^{2\pi i(ul+vm)} du dv.
\end{equation}

\noindent The coordinate space in which the measurements of $\V(u,v)$ are made is the $(u,v)$ plane or the aperture plane. We refer the reader to standard texts for the derivation of these relationships and to understand the various assumptions that underpin them (e.g., \citealt{Thompson1986, Perley1989} and later editions).  

In practice, a given antenna in a Fourier synthesis array collects radiation from the celestial source and couples it to the antenna electronics  through the antenna feed. The feed is generally designed to respond to radiation in a specific frequency range and to separate orthogonal senses of polarization -- either two linearly polarized signals X and Y, or two circularly polarized signals R and L. The signals from antenna $j$ are then amplified, conditioned\footnote{We ignore the many details involved in manipulating the signal before correlation: multiple stages of amplification, frequency conversion, filtering, digitization, etc.}, and corrected for geometrical delay before being multiplied by those from another antenna $k$, and averaged for an integration time $\tau$ at some specific frequency $\nu$ and bandwidth $\Delta\nu$ to produce a complex visibility measurement $\V_{jk}(u,v)$. The device that performs the cross-multiplications (or cross-correlations) and averaging is the {\sl correlator}. We will also discuss {\sl total power} measurements, $Z_i$, where the signal for each antenna is multiplied by itself (auto-correlations) and averaged. Note that we ignore the fact that antenna-based calibration factors (complex gains) must be derived and applied to the measured cross-correlations and auto-correlations to convert them to physical units. Calibrated visibilities are typically expressed in Jansky units. 

A Fourier synthesis telescope is a machine for measuring the complex visibility function $\V(u,v)$. A radio imaging array comprising $n$ antennas has $n_b=n(n-1)/2$ distinct antenna pairs, or {\sl antenna baselines}. Each antenna baseline measures a (complex) value of $\V(u,v)$, or single spatial Fourier component, of the radio brightness distribution of a celestial source at a given time, frequency $\nu$, bandwidth $\Delta\nu$, and wave number (we ignore polarization and spectral line observations here). Since the left hand side of Equation~2 is real $\V(u,v)$ is Hermitian, and so $\V(-u,-v)=\V^\ast(u,v)$. In practice, given the finite number of antennas in an array, the sampling of $\V(u,v)$ is discrete and non-uniform; there will be gaps in the sampling of $\V(u,v)$.  Fourier inversion of the visibility measurements therefore yields an imperfect image of the radio brightness distribution of the source. If we represent the sampling function by $s(u,v)$ the resulting map is

\begin{equation}
\I_D(\theta_x,\theta_y)=\int\int \V(u,v) s(u,v) e^{2\pi i(u\theta_x+v\theta_y)} du dv,
\end{equation}

\noindent where $A_\circ$ has been absorbed into $I_D$, often referred to as the ``dirty map". The sampling function is the autocorrelation function of the spatial distribution of antennas. Suppose a point source of unit flux density is observed at the phase tracking center. Then the real part of the complex visibility $\V^R(u,v)=1$ and the imaginary part is $\V^I(u,v)=0$ and the resulting dirty map is just 

\begin{equation}
\B_D(\theta_x,\theta_y)=\int\int s(u,v) e^{2\pi i(u\theta_x+v\theta_y)} du dv,
\end{equation}

\noindent referred to as the {\sl point spread function} (PSF) or the ``dirty beam".  The dirty map is a convolution of the true radio brightness distribution with the PSF, or dirty beam $\B_D$: $\I_D(\theta_x,\theta_y)=\B_D(\theta_x,\theta_y) \star \I(\theta_x,\theta_y)$. The PSF is typically characterized by a narrow central lobe with a width $\sim\!\theta_{\rm INT}$ and sidelobes. The main lobe is usually well-characterized by an elliptical Gaussian, referred to as the ``clean beam". Powerful nonlinear deconvolution or image regularization techniques are used to remove the sidelobes of the dirty beam from the dirty map to estimate the ``true" radio brightness distribution $\I(\theta_x,\theta_y)$, given the incomplete sampling of $\V(u,v)$ in the presence of the additive noise. We denote the estimate of the true radio brightness distribution $\I_C(\theta_x,\theta_y)$. 

The resolution of the clean map is characterized by the clean beam $\Omega_{\rm bm}\sim \theta_{\rm INT}^2$. Deconvolved maps made at radio wavelengths are typically expressed in units of spectral flux density per beam $\I_C\Omega_{\rm bm}$. It is often more convenient and informative to quantify the map in terms of brightness temperature $T_b$ (Kelvin units), which is linearly related to $\I_C\Omega_{\rm bm}$ through 
 
 \begin{equation}
\I_C(\theta_x,\theta_y)\Omega_{\rm bm}=\frac{2k_BT_b(\theta_x,\theta_y)}{\lambda^2}\Omega_{\rm bm}. 
\end{equation}

\noindent Most general purpose interferometers operating at radio wavelengths are designed to maximize sensitivity to faint celestial sources, parameterized as the {\sl signal-to-noise ratio} SNR $ = \I_C(\theta_x,\theta_y)\Omega_{\rm bm}/\sigma_I=T_b(\theta_x,\theta_y)/\sigma_T$, where $\sigma_I$ and $\sigma_T$ are the \rms\ values for maps expressed in units of flux density per beam or in Kelvin. The dynamic range (DR) is the maximum SNR in the map and is often used as a metric for the faintest signal that can reliably measured in a map in the presence of the brightest signal. 

It is common practice to resample the measured visibilities and the sampling function onto a uniform grid in order to exploit the Fast Fourier Transform (FFT) algorithm. In addition, the visibilities and sampling function are often weighted to optimize aspects of the map  — e.g., to achieve diffraction-limited angular resolution or to optimize surface brightness sensitivity. While the FFT is the basis for most radio astronomical imaging, it is also possible simply to compute the inverse Fourier transform directly for any point on the sky within the nominal field of view, an approach we adopt here for heuristic reasons. Hence, we can write the dirty map for  an $n$ antenna array with $n_b=n(n-1)/2$ baselines as

\begin{equation}
\I_D(\theta_x,\theta_y)=\frac{1}{2n_b}\sum_{j=1}^{n} \sum_{k>j}^{n}[V_{jk}(u,v)e^{-i(\theta_x u+\theta_y v)}+V_{jk}^\ast(u,v)e^{+i(\theta_x u+\theta_y v)}].
\end{equation}

\noindent In principle, an array can comprise antennas that also measure the total power entering each antenna, $Z_i$, equivalent to a visibility measurement at $(u,v)=(0,0)$. Including this measurement for each antenna yields the more general expression \citep{Vivek1991}

\begin{eqnarray}
\I_D^\circ(\theta_x,\theta_y)&=&\frac{1}{n^2}\bigl\lbrace \sum_{i=1}^n Z_i + \sum_{j=1}^{n} \sum_{k>j}^{n}[V_{jk}(u,v)e^{-i(\theta_x u+\theta_y v)}+V_{jk}^\ast(u,v)e^{+i(\theta_x u+\theta_y v)}]\bigr\rbrace \nonumber \\
\end{eqnarray}

\noindent We refer to radio interferometric arrays that do not include total power measurements as ``correlation arrays" and those which do include such measurements as ``total power arrays". Inclusion of total power measurements by each antenna in an array is not practical for reasons we discuss in Section~3.2 and Section~4.2; nor are they necessary for imaging purposes. The total flux density of a source can be introduced into imaging as a prior measured by, for example, a single dish. As we shall see, however, the inclusion of total power measurements when evaluating the map \rms\ noise does have a significant impact.  

In what follows we consider Fourier synthesis mapping in the so-called ``snapshot" imaging regime, by which we mean the maximum integration time, $\tau$, during which the array geometry is effectively fixed. While there are some subtleties associated with selecting an appropriate value for $\tau$ (see, for example \citealt{Thompson1986, Bridle1989}) it is of order seconds to tens of seconds for current and planned radio arrays used for solar observations (Appendix A).  We discuss circumstances under which the assumption of snapshot imaging can be relaxed in Paper II.

\section{Noise in Radio Observations: Preliminary Considerations}\label{Noise}

At radio wavelengths the Rayleigh-Jeans approximation is valid and it is common to refer signals to equivalent temperatures. A power input may be expressed as $P=k_B T\Delta\nu$, where $k_B$ is Boltzmann's constant, $\Delta\nu$ is the frequency bandwidth of the signal, and $T$ is the equivalent temperature. It is convenient to distinguish between the additive noise contributed by a given antenna and the noise contributed by the celestial source observed by the antenna. The former is described in terms of the {\sl system temperature} $T_{\rm sys}$, which involves a number of contributions:

\begin{equation}
T_{\rm sys}=T_{\rm rx}+T_{\rm bg}+T_{\rm sky}+T_{\rm spill}+T_{\rm loss}+T_{\rm cal} 
\end{equation}

\noindent where terms on the right-hand side are the contributions from the antenna receiver, from the galactic or cosmic microwave background, from the atmosphere, from ground radiation scattering into the feed, from losses due to various electronic elements, and from possible contributions due to a calibration signal injected into system. $T_{\rm rx}$ is typically the largest contribution to $T_{\rm sys}$, but not always (e.g., at meter wavelengths $T_{\rm bg}$ dominates).  

The incremental contribution of the radio source to the system noise is characterized by the {\sl antenna temperature} $T_{\rm ant}=S A_e/2k_B$ where $A_e=\eta A$ is the effective area of the antenna,  $\eta$ is the aperture efficiency, $A$ is the geometrical area of the antenna, and $S$ is the total flux density incident on the antenna. A convenient figure of merit for the sensitivity of an antenna is the {\sl system equivalent flux density}, defined as SEFD$=T_{\rm sys}/K$ where $K=A_e/2k_B$. The SEFD is the flux density of a hypothetical (unpolarized) source that would double $T_{\rm sys}$.  A lower value of the SEFD indicates better sensitivity, implying improvements may be realized by reducing $T_{\rm sys}$ and/or increasing $A_e$. Although $T_{\rm sys}$ and $T_{\rm ant}$ are commonly used to describe system noise and source noise, it is convenient for our purposes to use $N={\rm SEFD}$ instead of $T_{\rm sys}$ to refer to the noise contribution from an antenna, and $S=T_{\rm ant}/K$ to refer to the incremental source contribution instead of $T_{\rm ant}$. $N$ is an attribute of an antenna and $S$ is an attribute of the source. 

For most sources studied by radio telescopes $S\ll N$ (i.e., $T_{\rm ant}\ll T_{\rm sys}$) and, as we shall see, the noise properties across an image are uniform and easily quantifiable. Implicit in this case is the assumption that a radio interferometer yields $n_b$ independent measurements of the complex visibility function so the SNR of a synthesis image scales as $\sqrt{n_b}$. \citet{Kulkarni1989} pointed out that this cannot be true, in general, given that there are only $n$ independent noise contributions $N$ and $n$ signal contributions $S$. It is not possible to produce $n_b$ independent quantities based on $2n$ measurements. There must be correlations between antenna baselines. The theory of self-noise has been developed in several papers including work by \citet{Crane1989, Anantha1989, Kulkarni1989, Anantha1991, Vivek1991}, and \citet{Dewey1994}. The most extensive treatment is that of \citet{Kulkarni1989}, who considered noise in radio synthesis images for sources of arbitrary strength observed by correlating interferometers with arbitrary numbers of elements. 

We now discuss noise in radio measurements. In developing an understanding of noise and self-noise in the sections below we refer to simple unpolarized sources observed at the phase tracking center by various numbers of antennas in limits where $S\ll N$ or $S\gg N$.  Since we consider regimes that are outside the norm of most radio telescopes, we develop some intuition by first considering noise in single-dish observations, a two-element interferometer, and then generalizing to an $n$-element interferometer. For simplicity we consider measurements of an unpolarized source. 

\subsection{Single Antenna}

A single dish with an effective area $A_e$ can be used as a {\sl total power radiometer} for continuum observations over some frequency bandwidth $\Delta\nu\ll\nu$ and integration time $\tau$. Consider an unresolved source (i.e., $\theta_S<\theta_{\rm SD}\sim \lambda/D$) with a flux density $S$ at the center of the single dish beam The signal {\sl rms} measured by a total power radiometer is given by the {\sl ideal radiometer equation}:

\begin{equation}
\sigma_{\rm D}=\frac{S+N}{M}=\frac{Z}{N},
\end{equation}

\noindent where $M=\sqrt{\Delta\nu\tau}$ (see, e.g., \citealt{Condon2016} for a derivation).  Single dish observations are intrinsically total power measurements. We ignore non-ideal and variable contributions to the input signal due to, for example, receiver gain variations. 

It is usually the case that the source signal $S$ is very weak compared to system noise $N$. With $S\ll N$ we have 

\begin{equation}
\sigma_{\rm SD}\approx\frac{N}{M}=\frac{2k_B T_{\rm sys}}{M A_e}
\end{equation}

\noindent and SNR $=S/\sigma_{\rm SD}= SM/N$. For fixed $M$ the SNR is increased by minimizing $N$; i.e., by maximizing $A_e$ and minimizing $T_{\rm sys}$. In the case where the source signal is strong compared to the system noise, $S\gg N$, and 

\begin{equation}
\sigma_{\rm SD}\approx\frac{S}{M}=\frac{2k_BT_{\rm ant}}{MA_e}
\end{equation} 

\noindent and SNR $=M$. In this case, $A_e$ and $T_{\rm sys}$ are irrelevant, a result that is well known. If a map is made using a single dish using rastering techniques, a grid of discrete pointings, or a focal plane array the on-source noise is given by Equation~~(11) and the off-source noise is given by Equation~10. 

Now consider the case where a strong, uniformly bright, and extended source with a brightness temperature $T_b$ is observed by the single dish; for example, a uniformly bright disk of diameter $\theta_S$. When $\theta_S< \theta_{\rm SD}$ the signal {\sl rms} on the source is given by Equation~11. Suppose $D$ is increased. Then $\theta_{\rm SD}$ decreases and beyond some value of $D$ we have $\theta_S>\theta_{\rm SD}$ and the antenna temperature saturates at $T_{\rm ant}=T_b$. A further increase in $D$ does not lead to a further increase in $T_{\rm ant}$. In this case, $\sigma_D$ decreases as $1/A_e$ in Equation~11. 

\subsection{Two-element Correlating Interferometer}

We now consider a two-antenna correlation interferometer, the basic building block of a Fourier synthesis array. Assume that each antenna has effective area $A_e$ and that they are separated by a distance $d$. The two antennas are assumed to have identical electronics and orientations so that $N$ is also the same for each antenna. The antenna baseline is sensitive to angular scales $\theta_{\rm INT}\sim \lambda/d$. Consider an observation of a source on the zenith with a total flux density $S$ at the phase-tracking center of the interferometer. In general, the baseline measures a visibility with a correlated flux density amplitude $S_C<S$.  The noise fluctuations in the correlated flux are independent of those in the total signal and must therefore be added in quadrature to the total. The signal {\sl rms} is therefore \citep{Crane1989} 

\begin{equation}
\sigma_2 = \frac{1}{\sqrt{2}M}\Biggl[S_C^2+(S+N)^2\Biggr]^{1/2}.
\end{equation}

\noindent We first consider a point source with a flux density $S$ observed at the phase center. Then $S_C=S$ and Equation~13 reduces to the well-known result 

\begin{equation}
\sigma_2 = \frac{1}{M}\Biggl[S^2+{S N} + \frac{N^2}{2}\Biggr]^{1/2}.
\end{equation}

\noindent For $S\ll N$ we have 

\begin{equation}
\sigma_2 = \frac{N}{\sqrt{2} M} 
\end{equation}

\noindent and SNR $=S/\sigma_2=\sqrt{2}MS/N$. Comparing Equation~14 with Equation~10, it is seen that if the single dish has an effective area $A_{\rm SD}=2 A_e$ - that is, the sum of the areas of the two antennas in the interferometer - Equation~10 becomes  $\sigma_{\rm SD}=N/2M$ and SNR$=2MS/N$. The sensitivity of a two-element interferometer to weak sources, all other things being equal, is a factor $\sqrt{2}$ {\sl less} than that of a single dish with an effective area equal to that of the two single antennas combined. This is due to the fact that information in the auto-correlations, or total powers, is not included. In the opposite limit $S\gg N$ the source flux dominates:

\begin{equation}
\sigma_2 = \frac{S}{M}
\end{equation}

\noindent and again SNR $=M$, which is independent of $A_e$ and $T_{\rm sys}$, the same as the single-dish case. 

The fact that the noise is correlated between antennas leads to an {\sl rms} that is a factor $\sqrt{2}$ {\sl larger} than it would be for uncorrelated noise. It is straightforward to show \citep{Vivek1991} that inclusion of total power measurements in Equation~6 yields 

\begin{equation}
\sigma_2 = \frac{1}{M}\Bigl(S+\frac{N}{2}\Bigr).
\end{equation}

\noindent In this case, when $S\ll N$, $\sigma_2=N/2M$, the same expression as for a single dish with the same area as the sum of the two antennas in the interferometer. When $S\gg N$, we again have $\sigma_2=S/M$ and $A_e$ is again irrelevant for an unresolved source. If we compute the direct Fourier transform of the flux density at the phase tracking center for the two element interferometer using Equation~5 we simply have $\I_D^\circ(0,0)=S$. Including the total power measurements per Equation~7, however, we find that 

\begin{equation}
\I_D(0,0)=\frac{S+N}{2} + \frac{S}{2} = S+\frac{N}{2}
\end{equation}

\noindent and we see that $\sigma_2=\I^\circ_D(0,0)/M$ and SNR$ =M$.


Let us now consider a strong extended source. Equation~12 indicates that regardless of the correlated component $S_C$, there is a contribution from $S+N$. In particular, as $S_C\rightarrow 0$ due, for example, to the source filling the antenna FOV with a source of uniform brightness and being effectively ``resolved out'' by this baseline, the \rms\ noise is 

\begin{equation}
\sigma_2 =\frac{S+N}{\sqrt{2}M}.
\end{equation}

\noindent When $S\ll N$ we again have Equation~14 but when when $S\gg N$,

\begin{equation}
\sigma_2=\frac{S}{\sqrt{2}M}.
\end{equation}

\noindent This is the same form as Equation~14 but with the source noise $S$ playing the role of $N$. Since $S$ is uncorrelated between antennas in this case, $\sigma_2$ is reduced by factor of $\sqrt{2}$. The situation is analogous to a system with ``hot receivers" and a system temperature $T_{\rm ant}$. For a very extended source of mean brightness temperature $T_b$ that fills the antenna field of view $T_{\rm ant}=T_b$ as was discussed in \S3.1. 

\subsection{Correlating Array of $n$ Antennas}

We now generalize to a correlating array comprising $n$ antennas. An analytical expression analogous to Equation~13 for the signal {\sl rms} cannot be given for the general case, as we discuss below. However,  returning to the case of a point source at the phase tracking center, it can be shown (Appendix B; see also \citealt{Anantha1991, Vivek1991}) that the source {\sl rms} is 

\begin{equation}
\sigma_n = \frac{1}{M}\Biggl[S^2+\frac{2SN}{n} + \frac{N^2}{n(n-1)}\Biggr]^{1/2}.
\end{equation}

\noindent When $n=2$, Equation~13 results. When $S\ll N$, with $n_b=n(n-1)/2$, Equation~20 reduces to 

\begin{equation}
\sigma_n = \frac{N}{M\sqrt{2n_b}}.
\end{equation}

\noindent As $n$ becomes large, $2n_b\approx n^2$ and so 

\begin{equation}
\sigma_n=\frac{N}{nM}=\frac{2k_B T_{sys}}{nA_e M}.
\end{equation}

\noindent We see that the noise approaches that of a single dish (Equation~10) with a total effective area $nA_e$, and SNR $=SMn/N$. For $S\gg N$, Equation~20 once again yields $\sigma_n=S/M$ and SNR $= M$, which is independent of the number antennas, their effective area, or their summed area. 

By including total powers the expression for the point source {\sl rms} is simply 

\begin{equation}
\sigma_n=\frac{1}{M}\Bigl( S+\frac{N}{n} \Bigr).
\end{equation}

\noindent Clearly, as the number of antennas $n$ increases, Equation~20 converges to Equation~23. Note that the impact of the source signal becomes comparable to the system noise under relatively modest conditions, when $nS\sim N$.  Both equations yield the same limits for $S\ll N$ and $S\gg N$ for large $n$. If Equation~7 is used to compute the flux density at the phase tracking center we have 

\begin{equation}
\I_D^\circ(0,0)=\frac{S+N}{n} + \frac{(n-1)S}{n}=S+\frac{N}{n}
\end{equation}

\noindent and so $\sigma_n=\I_D^\circ(0,0)/M$ and, again, SNR $ = M$ at the location of the point source. 

For an extended source we again suppose $S_C\rightarrow 0$ on all baselines. Summing over all $n_b$ baselines and using Equation~12 we find

\begin{equation}
\sigma_n= \frac{1}{M}\Biggl[\frac{(S+N)^2}{2n_b}\Biggr]^{1/2}=\frac{1}{M}\frac{S+N}{\sqrt{2n_b}}
\end{equation}

\noindent which, for one antenna baseline, yields Equation~18. It is seen that in the case of an extended source observed by an array with large $n$ we always have a contribution to the noise of $(S+N)/nM$ which, for $S>>N$,

\begin{equation}
\sigma_{n} \approx \frac{S}{nM} = \frac{2k_B T_{\rm ant}}{MnA_e},
\end{equation}

\noindent where $S$ again plays the role of $N$ in Equation~22 and the baseline noise is determined by ``hot receivers" and a system temperature of $T_{\rm ant}$. When an extended source of brightness $T_b$ is resolved by the antennas, $T_{\rm ant}$ is replaced by $T_b$ in Equation~26. A critical difference between a point source and an extended source is that for a point source all baselines are fully correlated ($S=S_C$) and so there is no uncorrelated contribution from the ``hot receivers", only instrumental noise $N$; otherwise, for extended sources, a uniform noise floor given by Equation~25 or Equation~26 is present. 

\section{Noise in Fourier Synthesis Maps at Radio Wavelengths}

We now consider the case of an arbitrary radio brightness distribution and noise both on the source and off the source; i.e., where no radio emission is otherwise present on the sky. We first consider correlation arrays, widely used in practice, and then consider (hypothetical) total power arrays.

\subsection{Correlation Array}

A correlation array is one that makes use of all correlations between antennas, but does not make use of total power measurements. The {\sl rms} at any point in the field of view for a correlating array can be formed from the square root of the image variance. The image variance for a correlating interferometer of $n$ antennas observing a source of arbitrary strength and brightness distribution has been derived by \citet{Kulkarni1989}.  The result, expressed in terms of $[n_b \times n_b]$ covariance matrices, is cumbersome. In general, the image {\sl rms} depends on the details of the source brightness distribution, its strength, and the interferometric array used to observe it. The distribution of self-noise across an image obtained with a correlating interferometer with $n$ antennas is not obvious {\sl a priori}; it must be calculated explicitly for the circumstances in question although the discussion in previous subsections provides guidance. 

\begin{figure}
\centerline{\includegraphics[width=\textwidth]{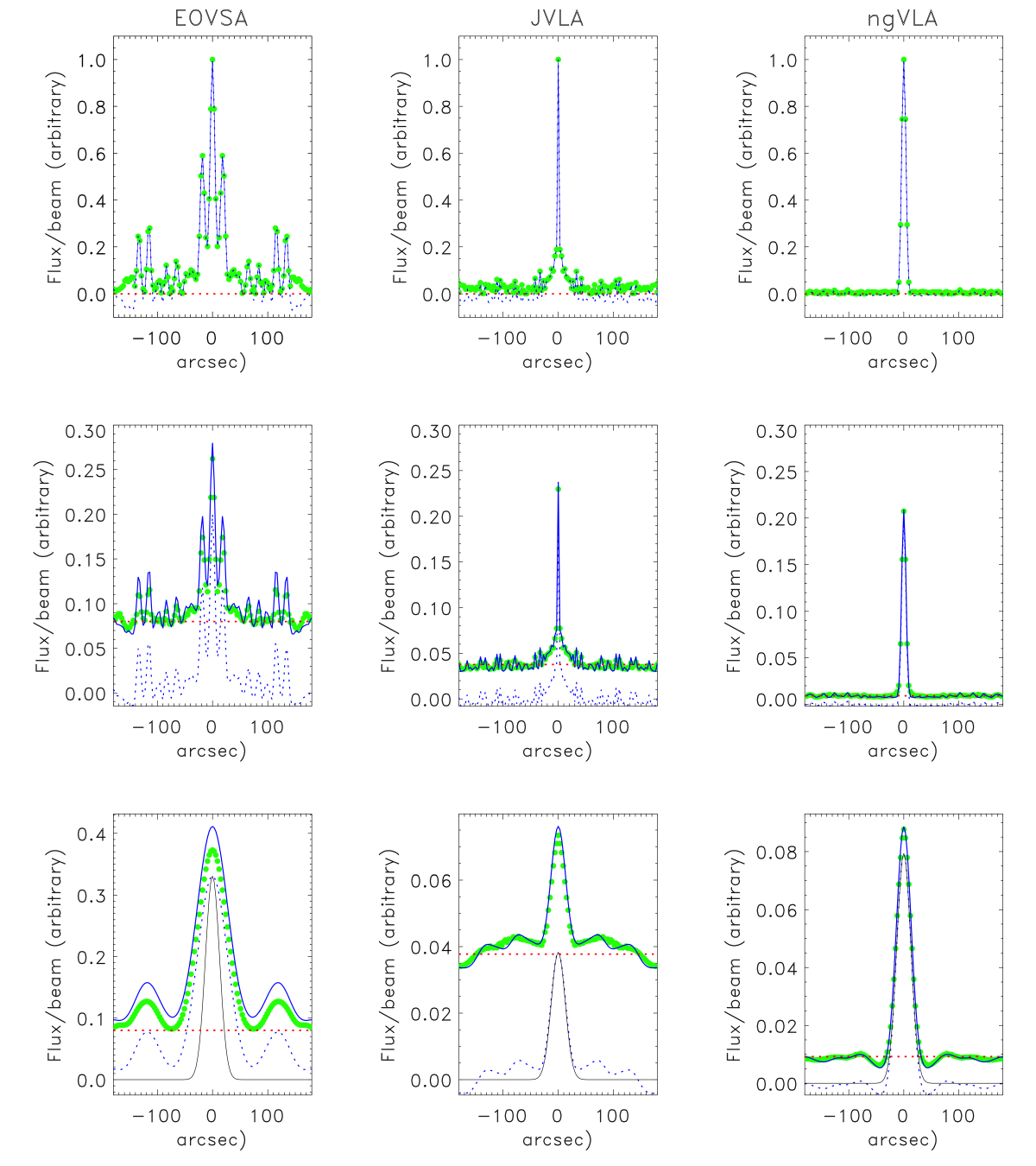}}
\caption{Comparison of model sources observed by EOVSA, the JVLA, and the ngVLA at a nominal frequency of 6~GHz. In each panel the {\it dashed blue line} represents $\I_D$, the {\it solid blue line} is $|\I_D+S/\sqrt{2n_b}|$, the {\it dashed red line} represents the noise floor, and the {\it green symbols} trace $\sigma_n$. Top row: the map and map {\sl rms} observed for a point source with $S=1$; Second row: the same for a point source with a flux density $S_{\rm pt}=0.2$ and a total flux $S_{\rm pt}+S_{\rm bg}=1$; Third row: The same for a Gaussian source with $\theta_G=30"$ and a total flux $S=1$. $N=0$ in all cases. Note the differences in scale for the ordinate.  }
\end{figure}

To gain insight into the general case we consider the map {\sl rms} for three simple source models: a point source, a point source observed against a uniform background, and an extended source. For the extended source we consider a simple symmetric Gaussian for which the expressions for the covariance matrix elements greatly simplify (Appendix B). In particular, in anticipation of Paper II, we compute the map rms in 1D (East-West) for three arrays:  the {\sl Expanded Owens Valley Solar Array} (\citealt{Gary2018}; EOVSA\footnote{The array is being upgraded to a 15-element array (expected to complete in 2026), which will have an improved PSF performance. The analysis presented here is based on the current 13-element array.}; $n=13$), the {\sl Jansky Very Large Array} in its C configuration (\citealt{Perley2011}; JVLA; $n=27$), and the core of the {\sl next generation Very Large Array} (\citealt{Murphy2018}; ngVLA; $n=114$). EOVSA is a solar-dedicated instrument that observes the Sun from 1-18 GHz ($\lambda 1.67-30$~cm), and the JVLA is a general purpose instrument capable of observing the Sun, also from 1-18 GHz. We include the proposed ngVLA because the antenna configuration of the core antenna distribution is relatively well established and it serves to illustrate the advantages of a large-$n$ array. It will be capable of observing the Sun from 1.2-116~GHz. The JVLA~C configuration and the ngVLA core have comparable sizes, with maximum antenna baselines of approximately 3~km. EOVSA, has a maximum baseline of approximately 1.2~km. We discuss these telescopes further in Paper II.

We adopt a representative frequency $\nu=6$~GHz ($\lambda=5$~cm) for all examples given. We take the total flux density to be $S=1$ unit and $N=0$. Results are shown in Figure~1. The top row shows the observation of a strong point source by each of the three arrays. The dotted blue line indicates the source map $\I_D(\theta_x)$ computed using Equation~6 and the solid blue line shows its absolute value. The green symbols show the corresponding map {\sl rms} $\sigma_n$, computed as the square root of the variance, scaled by $M$. The dashed red line indicates the zero-level in the top row. Since all baselines are fully correlated for a point source, there is no uncorrelated ``hot receiver" contribution to the noise as we discussed in Section~3.2. 

The second row shows the result for a point source of flux density $S_{\rm pt}=0.2$ on a uniform background contributing $S_{bg}=0.8$, their sum being $S=S_{\rm pt}+S_{\rm bg}=1$. The dashed blue line is again $\I_D(\theta_x)$, the solid blue line represents the $|\I_D+S/\sqrt{2n_b}|$, the green symbols show the scaled \rms, and the red dashed line is $S/\sqrt{2n_b}=1/\sqrt{2n_b}$. The correlated flux manifests approximately as the scaled PSF, but there is also an uncorrelated self-noise component (``hot receiver") that is added to the correlated component; that is, $\sigma_n(\theta_x)\approx [S_{\rm pt} \star PSF+(S_{\rm pt}+S_{\rm bg})/\sqrt{2n_b}]/M$. Careful inspection of the EOVSA \rms\ (green symbols) shows that they do not match the solid blue line, differing by $9\%$ at the location of the source. The difference at the source location is $4\%$ for the JVLA and $0.7\%$ for the ngVLA. Note that the same result would be obtained if we instead set the uniform background flux to zero ($S_{\rm bg}=0$) and let $N=0.8$. That is, a noise floor is always present as the result of uncorrelated contributions from $S$ and/or $N$. For both of the above examples, the map {\sl rms} is {\sl not} uniform across the FOV. The on-source noise is $\approx S/M$ and SNR$ \approx M$. The off-source noise is approximately proportional to the PSF and, therefore, the source sidelobes. Hence, for a point source observed against a background of uniform brightness, the correlated self-noise can be approximately removed from the map by subtracting the scaled PSF, leaving uniform noise plus a residual that decreases with $n$.

The third row shows results for a symmetric Gaussian source observed by each array. The full width at half maximum of the model Gaussian source is $\theta_G=30"$ and the integrated, or total, flux in the source is again 1~unit. The ordinate of each of the three plots varies because the source is resolved to varying degrees by each array and the peak flux density per beam therefore varies between arrays, particularly between EOVSA and the JVLA/ngVLA. The lines are the same as those in the top and middle rows. The black line is the model source. In this case, the difference between the green symbols and the solid blue lines is even more striking for EOVSA, differing by $11.5\%$ at the source maximum. It is $4\%$ and $0.7\%$ for the JVLA and ngVLA, respectively. Note that while the EOVSA PSF has a relatively narrow central lobe, it has large inner sidelobes as a result of sparse and non-optimum sampling of the {\sl uv} plane. The convolution of the PSF with the Gaussian model leads to a dirty map in which the source is apparently much broader than it is in reality. 

\begin{figure}
\centerline{\includegraphics[width=\textwidth]{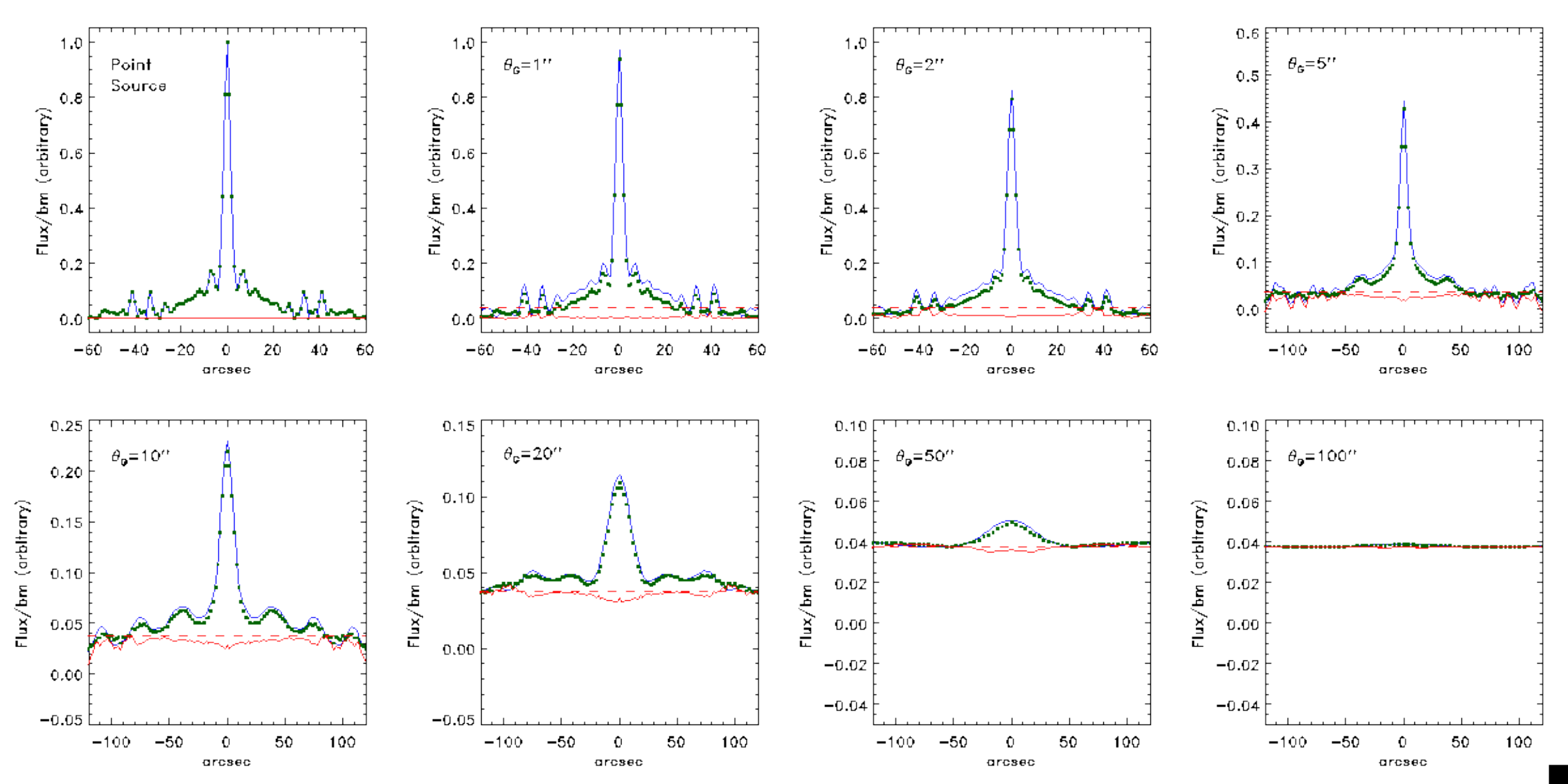}}
\caption{Comparison of model sources observed by the JVLA at a nominal frequency of 6~GHz. The source size varies from a point source to one that is well-resolved. All sources have $S=1$ and $N=0$. Notice that the off-source {\sl rms} rapidly approaches the noise floor given by Equation~26 as the source is resolved. The different lines are explained in the text. Note the differences in scale for both the ordinate and abcissa as the source size increases and the map flux per beam decreases.}
\end{figure}

\noindent While the self-noise contribution of a point source to off-source regions can be completely removed through deconvolution, such is not the case for strong extended sources.  As $n$ increases the map \rms\ increasingly resembles $\I_D$ and the off-source noise can be mitigated to the extent that sidelobes can be deconvolved from the map. Nevertheless, a noise floor will always exist, rapidly approaching $\approx S/\sqrt{2n_b}M$ as the source is resolved and $S$ becomes large.  Figure~2 illustrates this transition explicitly for a 6~GHz source observed by the JVLA, for which the angular resolution is nominally $3.4"$. We have calculated the source map and \rms\ in 1D for a point source and for Gaussians with FWHM values of $1"$, $2"$, $5"$, $10"$, $20"$, $50"$, and $100"$. All sources are again such that $S=1$ and $N=0$. The lines plotted again correspond to those shown in Figure~1 with the exception of the solid red lines. These show the difference between the computed \rms\ and the absolute value of the dirty map; i.e., $\sigma_I-|\I_D|$, which represents an estimate of the residual in the clean map (not including the scaling factor $M$). 

We have shown that as the number of antennas increases the calculated \rms\ for an extended source rapidly approaches the dirty map plus the noise floor, scaled by $M$. We therefore conclude that as $n$ becomes large we have for extended sources 

\begin{equation}
\sigma_I(\theta_x,\theta_y)\approx \frac{1}{M} \Bigl[ \I_D(\theta_x,\theta_y)+\frac{S+N}{\sqrt{2n_b}} \Bigr].
\end{equation}

In summary, for extended sources observed with correlation arrays, the map \rms\ increasingly resembles $\I_D$ plus a uniform noise component that depends on both $S$ and $N$ as $n$ increase. While off-source self-noise due to source sidelobes can be mitigated through deconvolution to an increasing degree with increasing $n$, it cannot entirely removed. Moreover, a noise floor is always present as a result of uncorrelated contributions from the total source flux $S$ and instrumental noise $N$. The noise floor decreases with increasing $n$ but it cannot be removed through deconvolution. 

\subsection{Total Power Array}

From the discussion of single dish observations in Section~3.1, we expect a uniform {\sl rms} $\sigma_{\rm SD}=N/M=2k_B T_{\rm sys}/M A_{\rm SD}$ in regions free of strong emission and we would expect self-noise with $\sigma_{\rm SD}=S/M$ to be relevant only at locations on the strong source. Consider a correlation array  where the sum of the antenna areas is the same as that of the single dish: $A_{\rm SD}=nA_e$. We would expect the off-source noise to be $\sigma_n=\sigma_{\rm SD}=N/M=2k_BT_{\rm sys}/MnA_e$, but we found in that it is instead $\sigma_n=S/nM=2k_BT_{\rm ant}/MnA_e$ (Equation~25). How do we reconcile expectations from single-dish mapping with Fourier synthesis mapping of strong sources with a correlation array? 

Single dish radiometers make total power measurements but correlation arrays only use cross-correlations between antennas and they do not include total power measurements. \citet{Vivek1991} and \citet{Anantha1991} show, however, that the inclusion of total power measurements in calculating the covariance matrix greatly simplifies the expression for the image variance given by \citet{Kulkarni1989} to an elegant and intuitive form, one that we hinted in Sections~3.2 and 3.3 (Equations~17 and 24). Denoting the total power for antenna $i$ as $Z_i=S_i+N_i$ the inclusion of covariance terms between $Z_i$ and $Z_i$, $Z_i$ and $Z_j$, of $Z_i$ with all baselines containing antenna $i$, and of $Z_i$ with all baselines not involving antenna $i$ allows a square to be completed in the sum over all covariances. The result for the \rms\ anywhere in the map is just 

\begin{equation}
\sigma_I (\theta_x,\theta_y)=\frac{1}{M} \Bigl[I^\circ_D(\theta_x,\theta_y)+\frac{N} {n}\Bigr].
\end{equation}

This raises a subtle and important distinction. For a total power array, the self-noise in the total flux $S$ measured by each antennas is fully correlated with that of every other antenna while the system noise $N$ of each antenna remains uncorrelated between antennas. In this case, the total flux $S$ can be subsumed into the dirty map and a {\sl uv} sample at $(u,v)=(0,0)$ is included in the PSF; i.e., we have $V(0,0)=S$ contributing to the map. Under these circumstances, the sidelobes and self-noise may be removed completely from off-source locations through deconvolution, leaving only the uncorrelated system noise $N/nM$, in agreement with the single dish result. This being the case, why aren't total power arrays the norm? There are two reasons: first, there is no particular advantage in producing a synthesis map made with a total power array compared with a correlating array. Total power measurements can be introduced to image reconstruction as a prior, if necessary, as discussed in Section~2. Second, for most modern radio astronomical observations $S<<N$ and self-noise plays no significant role in a Fourier synthesis map made with either a total power array or a correlation array. The map \rms\ is uniform with $\sigma_n=N/M\sqrt{2n_b}\approx N/Mn$. A total power array is a factor $\sqrt{n/(n-1)}$ more sensitive than a correlation array, which is 8\%, 2\%, and 0.4\% for EOVSA, the JVLA, and the ngVLA core, respectively. The penalty in sensitivity incurred by neglecting total power measurements rapidly diminishes with increasing $n$. The expense of total power measurements with each antenna is not justified, in general, especially for large-$n$ arrays. It is for these reasons that no total power arrays have been implemented in practice. It must be acknowledged, however, that observations of strong extended sources pay a significant noise penalty because $S/nM>>N/nM$. We explore the consequences of this in Paper II for solar observations. 

\section{Concluding Remarks}

We have outlined the theory of self-noise in Fourier synthesis maps at radio wavelengths in the snapshot imaging regime. We first considered observations of simple sources — point sources or very extended sources — using a single dish, a two-element interferometer, and an $n$-element array to develop intuition. We then considered the case of an arbitrary brightness distribution of arbitrary strength and the noise both on and off the source. For a correlation array, one for which total power measurements are not included, the map \rms\ must be computed explicitly, in general. For point sources, the map {\sl rms} is just the scaled PSF; for a point source observed against a uniform background, the map {\sl rms} is the scaled PSF plus an offset due to uncorrelated noise (instrumental plus self-noise). For an extended source, illustrated by a Gaussian, the map \rms\ has a complex distribution for a small-$n$ array like EOVSA. While the dirty map $\I_D$ resembles the map \rms, the two differ in significant ways. However, as $n$ becomes large, the map \rms\ approaches $\I_D$, offset by the uncorrelated noise floor and scaled by $M=\sqrt{\Delta\nu\tau}$. Off-source self-noise can therefore be mitigated to an increasing degree through deconvolution as $n$ increases, but the uniform noise floor due to uncorrelated source and instrumental noise cannot be removed. We compare this result with a total power array, and $n$-element correlating array that includes total power measurements. In this case, the map \rms\ is strictly proportional to $\I_D^\circ$ plus uniform instrumental noise, even for extended sources. In this case, the dirty map $\I_D^\circ$ includes the total source flux $\V(0,0)=S$. The self-noise contribution can be completely removed from snapshot maps through deconvolution, leaving only a uniform contribution due to uncorrelated instrumental noise. In practice, however, the expense of implementing total power arrays is not justified for the vast majority of observations of celestial sources. 

In Paper II, we use these results to consider limitations imposed by self-noise on solar observations at radio wavelengths on current and planned radio Fourier synthesis imaging arrays. Our intent is to understand the degree to which self-noise is a limiting factor for various science use cases, with the aim of developing instrument requirements and imaging strategies that minimize its impact.  

\begin{acks}
The National Radio Astronomy Observatory is a facility of the National Science Foundation operated under cooperative agreement by Associated Universities, Inc. The Expanded Owens Valley Solar Array (EOVSA) was designed, built, and is now operated by the New Jersey Institute of Technology (NJIT) as a community facility.
\end{acks}

\begin{fundinginformation}
 EOVSA operations are supported by NSF grant AGS-2436999 and NASA grant 80NSSC20K0026 to NJIT.
\end{fundinginformation}

\newpage

\appendix

\section{Snapshot Imaging Regime}

In \S2 the snapshot imaging regime was described as that in which the array geometry remained effectively fixed for an integration time $\tau$. There are subtleties, however. The snapshot integration time also depends on the frequency bandwidth, $\Delta\nu$, and the size of the domain being imaged. A given antenna baseline in an array traces out an elliptical track in the {\sl uv} plane with time due to the Earth's rotation, a fact exploited by Earth rotation aperture synthesis. The track is quasi-azimuthal. A given {\sl uv} point also changes radially with frequency, which can be exploited for the purposes of multi-frequency synthesis. Both Earth rotation aperture synthesis and multi-frequency synthesis are discussed further in Paper II. An integration time that is too long results in temporal smearing of the visibility measurement on a given baseline. Similarly, the use of too large a frequency bandwidth results in smearing in the radial direction. 

Consider once again a point source. It suffers no radial smearing at the phase tracking center but it suffers an increasing degree of bandwidth smearing as its angular distance from the phase tracking center increases. As a result, the amplitude of the point source is reduced. It is straightforward to estimate the maximum bandwidth over which the visibility data can be averaged to minimize the bandwidth smearing over the imaging FOV \citep{Bridle1989}. For example, to ensure an amplitude loss of no more than 1\% when imaging the full disk of the Sun with an array of 2~m antennas — that is, no more than 1\% at the limb — the bandwidth should be no more than $\Delta\nu=4.4$~MHz at any wavelength. For an array of 25~m antennas, with their wavelength-dependent FOV, we require as reduction, say, of no more than 1\% in amplitude at the half-power point of the FOV. The bandwidth is then $\Delta\nu\approx 90/\lambda$~MHz, yielding $\Delta\nu\approx 3$~MHz at 1~GHz and $\delta\nu\approx 60$~MHz at 20~GHz. This is not to say that a large effective bandwidth cannot be created for multi-frequency synthesis, but the net bandwidth must be channelized such that each frequency channel is no larger than these values. 
 
Similar to bandwidth smearing in the radial direction is quasi-azimuthal smearing in the image domain as the result of time averaging. Time-average smearing also increases with angular distance from the phase-tracking center. To ensure that time-average smearing is no worse than bandwidth smearing - that both are $1\%$ or less over the angular domain being imaged, for example - we take $\tau\approx \Delta\nu/\nu\omega_e$ where $\omega_e=7.27\times 10^{-5}$ is the angular rate of Earth's rotation. For an array of 2~m antennas observing the full disk of the Sun with no more than $1\%$ smearing across the source due to finite bandwidth or time averaging, we have $\tau_a=60$~s at 1~GHz and 3~s at 20~GHz; for an array of 25~m antennas imaging the Sun, we have $\tau_a \approx 40$~s at all wavelengths. 

The values given here for $\tau$ and $\Delta\nu$ are illustrative. Specific snapshot imaging strategies depend on the details of the array and the science use case in question. 

\section{Image Variance of a Strong Symmetric Source}

The complete and general expression for the variance of a Fourier synthesized image at radio wavelengths for an array comprising $n$ antennas is given by \citet{Kulkarni1989}:

\begin{align}
\sigma_I^2(\theta_x,\theta_y)= \frac{1}{n_b^2} \sum_{j=1}^n \sum_{k>j}^n \sum_{j'=1}^n \sum_{k'>j'}^n \Bigl\lbrace&a_{jk}(\theta_x,\theta_y) a_{j'k'}(\theta_x,\theta_y)C[v_{jk}^c,v_{j'k'}^c] \nonumber \\
+&a_{jk}(\theta_x,\theta_y) b_{j'k'}(\theta_x,\theta_y)C[v_{jk}^c,v_{j'k'}^s] \nonumber \\
+&b_{jk}(\theta_x,\theta_y) a_{j'k'}(\theta_x,\theta_y)C[v_{jk}^s,v_{j'k'}^c] \nonumber \\
+&b_{jk}(\theta_x,\theta_y) b_{j'k'}(\theta_x,\theta_y)C[v_{jk}^s,v_{j'k'}^s] \nonumber \Bigr\rbrace
\end{align}

\noindent where $C[v_{jk}^c,v_{j'k'}^c$, $C[v_{jk}^c,v_{j'k'}^s]$, $C[v_{jk}^s,v_{j'k'}^c]$, and $C[v_{jk}^s,v_{j'k'}^s]$ are $[n_b\times n_b]$ covariance matrices of correlations between antenna baselines and 

\begin{align}
a_{jk}(\theta_x,\theta_y)&=\cos[2\pi(u\theta_x+v\theta_y)] \nonumber \\
b_{jk}(\theta_x,\theta_y)&=\sin[2\pi(u\theta_x+v\theta_y)] \nonumber
\end{align}

\noindent where $u$ and $v$ are defined in \S2. To simplify the discussion while retaining key insights into the nature of source variance for strong signals, we assume that the source is symmetric and centered on $(\theta_x,\theta_y)=(0,0)$. An example is a circular Gaussian with a full width at half maximum of $\theta_G$, as discussed in \S3.4. Since the source is symmetric about the phase-tracking center the real part of the visibility is non-zero and the imaginary part is zero. We denote the visibility on baseline {\sl jk} as $V_{jk}$. Under these assumptions the image covariance becomes

\begin{align}
\sigma_I^2(\theta_x,\theta_y)= \frac{1}{n_b^2} \sum_{j=1}^n \sum_{k>j}^n \sum_{j'=1}^n \sum_{k'>j'}^n \Bigl\lbrace&a_{jk}(\theta_x,\theta_y) a_{j'k'}(\theta_x,\theta_y)C[v_{jk}^c,v_{j'k'}^c] \nonumber \\
+&b_{jk}(\theta_x,\theta_y) b_{j'k'}(\theta_x,\theta_y)C[v_{jk}^s,v_{j'k'}^s] \nonumber \Bigr\rbrace
\end{align}

\noindent The expression for the covariance matrix elements depends on their type: the diagonal elements comprise covariances between like baselines -- or simply, variances (e.g., $b_{12}$ with $b_{12}$ - type $a$); off-diagonal elements that share a common antenna (e.g., $b_{12}$ with $b_{13}$ - type $b$); and off-diagonal elements involving baselines that share no antennas (e.g., baseline $b_{12}$ with $b_{34}$ - type $c$).  The expressions for the diagonal elements (type a), of which there are $n_b$, are:

\begin{equation}
C[v^c_{jk},v^c_{jk}]=\frac{(S+N)^2+V_{jk}^2}{2M^2}; \qquad C[v^s_{jk},v^s_{jk}]=\frac{(S+N)^2-V_{jk}^2}{2M^2} \nonumber
\end{equation}

\noindent The covariances of off-diagonal elements of type $b$, of which there are $2n_b(n-2)$, are 

\begin{equation}
C[v^c_{jk},v^c_{jl}]=\frac{(S+N)V_{kl}+V_{jl}V_{jk}}{2M^2}; \qquad C[v^s_{jk},v^s_{jl}]=\frac{(S+N)V_{kl}-V_{jl}V_{jk}}{2M^2} \nonumber
\end{equation}

\noindent and the covariances of off-diagonal elements of type $c$, of which there are $n_b(n-2)(n-3)/2$, are

\begin{equation}
C[v^c_{jk},v^c_{lm}]=\frac{V_{jl}V_{km}+V_{jm}V_{kl}}{2M^2}; \qquad
C[v^s_{jk},v^s_{lm}]=\frac{V_{jl}V_{km}-V_{jm}V_{kl}}{2M^2} \nonumber
 \end{equation}

\noindent The image variance is straightforward to calculate for specific arrays but since the number of terms increases as $n_{bl}^2\sim n^4$ it can become computationally intensive for large-$n$ arrays. Note that for a point source with a flux density $S$, we have $V_{jk}=S$ for all baselines and so

\begin{eqnarray}
C[v^c_{jk},v^c_{jk}]=[(S+N)^2+S^2]/2M^2&;& \qquad C[v^s_{jk},v^s_{jk}]=[(S+N)^2-S^2]/2M^2 \nonumber \\
C[v^c_{jk},v^c_{jl}]=[(S+N)S+S^2]/2M^2&;& \qquad C[v^s_{jk},v^s_{jl}]=[(S+N)S-S^2]/2M^2 \nonumber \\
C[v^c_{jk},v^c_{lm}]=2S^2/2M^2&;& \qquad C[v^s_{jk},v^s_{lm}]=0 \nonumber
\end{eqnarray}

\noindent for matrix elements of type $a$, $b$, and $c$, respectively, from which Equation~21 follows.

\newpage

\bibliography{Noise.bib}{}
\bibliographystyle{aasjournal}

\end{article}
\end{document}